\documentclass[doublecol]{epl2}

\usepackage{amsmath}
\usepackage{amsfonts}
\usepackage{amssymb}
\usepackage{graphicx}
\usepackage{dcolumn}
\usepackage{bm}
\usepackage{textcomp}
\usepackage{footnote}

\title{Oscillatory Shannon Entropy in the Process of Equilibration of Nonequilibrium Systems}
\author{ A. Giri \and Nilangshu K. Das \and P. Barat\footnote{ Corresponding author : pbarat@vecc.gov.in (P. Barat), Tel: +91 03323593467, Fax: +91 03323346871 }  }
\shortauthor{A. Giri \etal}
\institute{Variable Energy Cyclotron Centre, 1/AF Bidhannagar, Kolkata 700064, India}

\pacs {89.70.Cf} {Entropy and other measures of information} 
\pacs {65.40.gd}{Entropy in condensed matter}
\pacs{05.70.Ln}{Nonequilibrium and irreversible thermodynamics}

\abstract
    {We present a study of the equilibration process of nonequilibrium  systems by means of molecular dynamics simulation technique. The  nonequilibrium conditions are achieved in systems by defining velocity components of the constituent atoms randomly. The calculated Shannon entropy from the probability distribution of the kinetic energy among the atoms at different instants during the process of equilibration shows oscillation as the system relaxes towards equilibrium. Fourier transformations of these oscillating Shannon entropies reveal the existance of Debye frequency of the concerned system. From these studies it was concluded that the signature of the equilibration process of dynamical systems is the time invariance of Shannon entropy.}

\begin{document}
\maketitle

Most of the systems observed in nature or in the laboratories are in the nonequilibrium state. Physical properties of systems are measured by bringing them in their nonequilibrium state by applying external peturbations. In last two decades there was lot of impetus to understand the process of equilibration of nonequilibrium system, far from equilibrium, by studying the dynamical evolution of the individual constituents of the system \cite{Seifert}. Understanding of the collective behaviour of nonequilibrium systems is poor thus, it is important to study the behaviour of the ensemble of atoms of a nonequilibrium system together during the passage of equilibration. This can be achieved by studying the variation of the probability distribution function (PDF) of the energy of the associated atoms in the system during equilibration. The global dynamics of equilibrium systems are studied by statistical mechanics forging the fundamental link between the interactions of the constituents and the macroscopic behaviour of the interacting many body systems. Boltzmann established a general framework to evaluate the associate probability of an ensemble of an equilibrium system to achieve a particular energy state. This simple relation establishes several macroscopic thermodynamic parameters of the equilibrium system. Boltzmann distribution is very rigid and derived on the concept of maximum thermodynamic entropy (Boltzmann entropy) or from the most probable distribution. Thus, whenever an nonequilibrium system reaches an equilibrium state it has to follow the path of steepest thermodynamic entropy ascent compatible with the constraints of the system concerned. This observation is in consistant with Onsager theory of reciprocity \cite{Giovanni} and fluctuation-dissipation theory \cite{Kubo}. 

A nonequilibrium system continuously undergoes transitions from one state to the other to maximize the Boltzmann entropy and at equilibrium the distribution reaches the Boltzmann distribution. Thus it is imperative to study the evolution process of equilibration of a nonequilibrium system one has to calculate the time dependent PDF, $p(E,t)$ and the corresponding measure of entropy. In this Letter we present the time variation of the Shannon entropy \cite{Shannon} of nonequilibrium systems defined by the relation  
$S(t) = - \int p(E,t) \, log \, p(E,t) \,dE,$ 
where $p(E,t)$ is obtained by molecular dynamics (MD) simulation technique.

A variety of materials having different crystal structures with widely employed emperical interaction potentials are used to carry out the MD simulations \cite{Pbarat}. Depending on the potential function, the trajectories of the atoms are calculated at each time step of simulation by solving Newton's laws of motion. Copper (Cu), aluminium (Al) and solid argon (Ar) with face centered cubic (FCC) structure, and iron (Fe) with body centered cubic (BCC) structure are the elements considered in this study. The simulations are performed using MD++ \cite{MD} code. The emperical potentials used are Embedded Atom Method, Aluminum Glue, Lennard Jones and Finnis-Sinclair for Cu, Al, solid Ar and Fe respectively. Cubic cells of size 20\texttimes20\texttimes20 unit cell (uc) containing 32000 Cu, Al and solid Ar atoms, and  25\texttimes25\texttimes25 uc containing 31250 Fe atoms are considered as the simulation systems. Number of atoms in our simulations are much larger than that recommended in reference \cite{sand} to avoid the size effect. In all the cases, initially, the systems are relaxed at 100K (30K for solid Ar) temperature for 2.5ps (10ps for solid Ar ) using periodic boundary condition in all three directions under constant number of atoms, volume and total energy (NVE) ensemble such that the total energy of the system, according to equipartition theorem, gets distributed equally making the average kinetic energy (KE) and the average potential energy (PE) equal. The number of time steps used for relaxations was 5000 with each time step of 0.5fs (2fs for solid Ar).
\begin{figure}[h!]
\includegraphics[width= 0.45\textwidth]{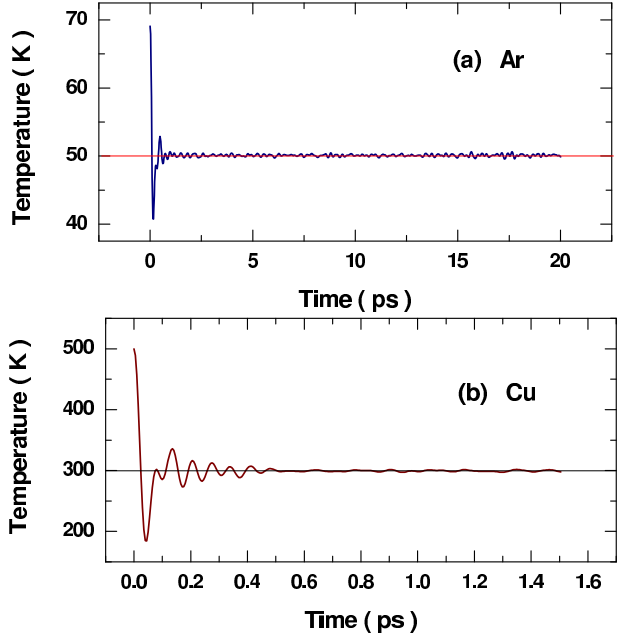}
\vspace{-10pt}
\caption{Variations of the average temperature (equivalent to average KE ) with time of the systems of (a) Ar and (b) Cu leading to the equilibration.}
\vspace{-5pt}
\label{figg1}
\end{figure}
To set up the nonequilibrium state the  velocity  components of the atoms are redefined such that the average KE of the atoms becomes 500K (70K for solid Ar). The systems are then released for equilibration. Consequently the systems gradually proceed towards equilibrium with the average temperature approaching to 300K (50K for solid Ar) as shown in Fig.~\ref{figg1}
 
\begin{figure}[h!]
\includegraphics[width= 0.46\textwidth]{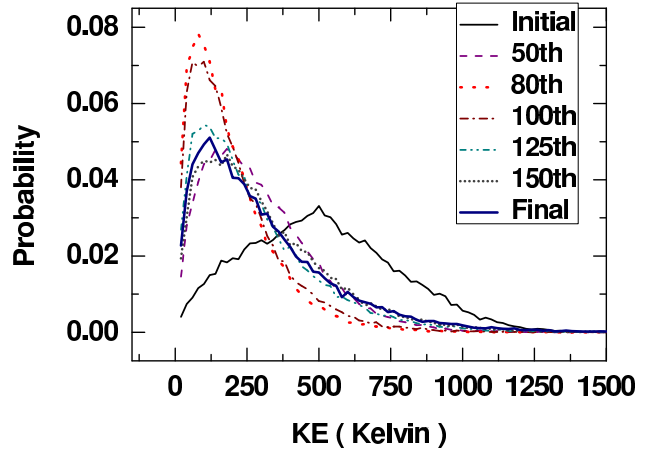}
\vspace{-10pt}
\caption{The probability distribution function of the kinetic energy (expressed in terms of temperature) of Cu atoms at initial, final and at five different time steps.}
\vspace{-5pt}
\label{figg2}
\end{figure}

\begin{figure}[h!]
\includegraphics[width= 0.45\textwidth]{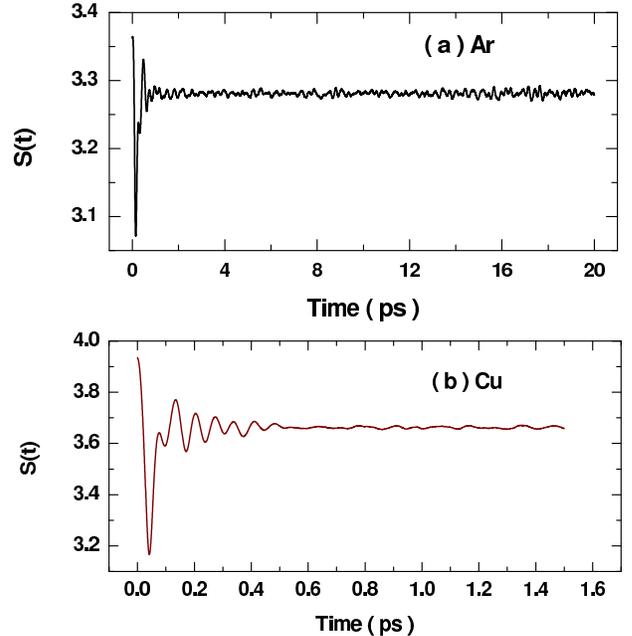}
\vspace{-10pt}
\caption{Variation of the Shannon entropy with time of (a) Ar and (b) Cu during the course of equilibration.}
\vspace{-5pt}
\label{figg3}
\end{figure}

 During the equilibration process the position and velocity components of each atom are recorded at every time steps of 0.5fs (2fs for solid Ar). The total equilibration time was 1.5ps (20ps for solid Ar). At each timestep the KE spectrum of the atoms are divided into 200 bins (130 bins for slid Ar) with each bin corresponds to KE of 20K (5K for solid Ar) in oder to get the histogram of the KE spectrum. The normalized probability distribution of KE of Cu atoms expressed in terms of temperature during the process of  equilibration are shown in Fig.~\ref{figg2}. 
After making the normalisation
$\sum _{i=1}^N p_{i}(t)=1$ 
( N = number of bins) of the KE spectrum the values of  the Shannon entropy 
$S(t)=-\sum _{i=1}^N p_{i}(t)logp_{i}(t)$
at different instants of time are calculated. The variations of S(t) with time for Ar and Cu are shown in Fig.~\ref{figg3}

\begin{table*}[t]
\caption{ Frequencies of the first peak observed from the Fast Fourier Transform of Shannon entropy of the elements and their respective Debye frequencies.}
\label{table2}
\begin{center}
\begin{tabular}{lccc}
\hline
\hline
\\
\bf{Material} & \bf{1st peak from FFT}  & \bf{Debye frequency} \\\\
\hline
\\

Solid Argon & $1.93  $ THz   & $1.77 $ THz\\
Copper & $7.14 $ THz & $7.16 $ THz \\
Iron & $9.74 $ THz & $9.79$ THz\\
Aluminium & $7.81 $ THz  & $8.21$ THz \\\\
\hline
\hline
\end{tabular}
\end{center}
\end{table*}

  Ensemble of atoms in a crystal when brought to nonequilibrium state by enhancing the KE of the atoms from their equilibrium values, equilibrate by diffusing its excess KE. This diffusion process is governed by the various forces acting on the atoms. The force arising from the nearest neighbour interaction potential plays the role of the external force on the atom. There will be energy exchange between the atom and surrounding atoms in a result of which the atom may lose a part of its KE. Loss of KE also occurs because of its conversion to PE in the ensemble of atoms. This can be looked as the effect of a frictional force acting on the atom. There will be a force acting on the atom arising from the interactions from all other atoms in the cell. Because of their thermal vibrations the net force will be random. Effect of these forces results in a change in the KE of the atoms and consequently in the PDF. The PDF changes with time and ultimately equilibrates to the stable Boltzmann distribution. The variation of the PDF of KE can be a measure of the dynamical process of equilibration. In this regard Shannon entropy $S(t)$ is evoked and it has been calculated at each timestep.

\begin{figure}[h!]
\includegraphics[width= 0.45\textwidth]{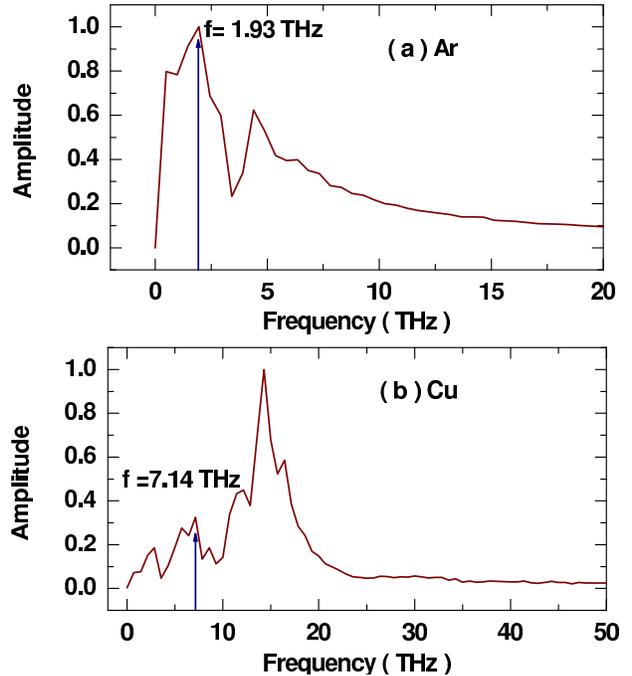}
\vspace{-10pt}
\caption{FFT of Shannon entropy of (a) Ar and (b) Cu.}
\vspace{-5pt}
\label{figg4}
\end{figure}

The Shannon entropy oscillates (Fig.~\ref{figg3}) around the equilibrium value during the process of equilibration. Thus high and low values of the Shannon entropy does not necessarily signify any stable state of the system but the invariance of the Shannon entropy with time suggests the attainment of equilibrium. Thus in equilibrium of any dynamical system Shannon entropy is a constant of motion. Fast Fourier Transformation (FFT) of the time evolution of the Shannon entropy shows a frequency spectrum having two major peaks as shown in Fig.~\ref{figg4}. The first peak is closed to the Debye frequency of the element concerned as shown in Table~\ref{table2}.

In the study of the statistical interpretation of the relaxation process of nonequilibrium systems, Boltzmann entropy plays the most important role and it attains the maximum value at equilibrium. However, Boltzmann entropy can not be estimated based on the concept of probabilities as defined in the field of statistics and as well difficult to measure in numerical studies. Thermodynamic probabilities or the statistical weight of a microstate that reflects the Boltzmann entropy is not a probability as defined in the conventional statistics \cite{Chakrabarti}. Hence, in our numerical studies of the relaxation process we are compelled to evoke the Shannon entropy to understand the dynamics. Shannon entropy is a measure of uncertainty \cite{Shannon} and is a positive function of $p_i$s. Its extreme values are for the cases when all $p_i$s are equal with $S = log N$ and $S=0$ when the system is uniquely defined with sharped peak at one of the $p_i$s equal to 1.

When the system is far away from the equilibrium normalized values of probability in the respective bins are much different from the equilibrium values (Fig.~\ref{figg2}). When the system is allowed to relax the values of probabilities start oscillating in such a way that these coupled oscillators reach the final destination correspond to the Boltzmann distribution. The behavior of the PDF at diffrent time steps (Fig.~\ref{figg2}) shows that, initially the distribution starts peaking at the low temperature side in an attempt to approach the Boltzmann distribution. At 80th time step (for Cu) it peaked maximum and the corresponding Shannon entropy goes to minimum. However, as the coupled oscillating probabilities in the bins do not reach the equilibrium Boltzmann distribution it again starts flattening making the Shannon entropy to increase. This cumulative oscillation among the bins persists but its amplitude decays with time so long it reaches Boltzmann distribution making the Shannon entropy as a constant of motion. This is the global dynamics of these equilibrium processes.

In this microcanonical ensemble whenever the KE is exchanged between the atoms it generates a different microstate for a given macrostate. This energy exchange is also responsible for the oscillations of the probabilities in the bins. These oscillators when coupled together generate higher modes of oscillations along with their fundamental frequency. Because of these coupled oscillators we get the frequency spectrum of the time dependence of the Shannon entropy. We presume that the observed first peak in the frequency spectrum is the Debye frequency and is one of the primary frequencies responsible for the transfer of probabilities between the bins.

The nature of the variation of average temperature with time and the Shannon entropy as shown in Fig.~\ref{figg1} and Fig.~\ref{figg3} respectively have identical character. Initially the dip in the average temperature corresponds to the peaking of the PDF at low temperature side and consequently a downfall in the Shannon entropy observed. Reduction of average temperature signifies that the KE of the system transforms to the PE. This oscillation of energy exchange between KE and PE continues so long the average KE equals to the average PE. At that juncture the system attains the equilibrium.

One of the characteristics of the nonequilibrium systems is that the PDF associated with the systems are necessarily time dependent. Because of this time dependence it is difficult to find any general analytical formalism to deal with its dynamics. The time evolution of $p_i$s is governed by the master equation. However, it is extremely difficult to solve it numerically because of associated large degrees of freedom. In our attempt, we could justify the evolution dynamics of nonequilibrium systems by studing their time evolution of the Shannon entropy and we could conclude that the time variation is primarily dictated by two frequencies and one of them is the Debye frequency.



\begin{thebibliography}{99}

\bibitem{Seifert} Udo Seifert, {\it Rep. Prog. Phys.}, \textbf{75} (2012) 126001 and the references therein.
\bibitem{Giovanni} Giovanni Gallavotti, {\it Phys. Rev. Lett.}, \textbf{77} (1996) 4334.
\bibitem{Kubo} R. Kubo, {\it Rep. Prog. Phys.}, \textbf{ 29} (1966) 255 and the references therein.
\bibitem{Shannon}C. E. Shannon, {\it The Bell System Technical Journal}, \textbf{ 27} (1948) 379.
\bibitem{Pbarat}P. Barat, A. Giri, M. Bhattacharya, Nilangshu K. Das and A. Dutta, {\it Europhys. Lett.}, \textbf{104} (2013) 50003.
\bibitem{MD} http://micro.stanford.edu/caiwei/Forum/2004-12-12-MD++/.
\bibitem{sand} David Sands, Jeremy Dunning-Davies, arXiv:1301.1364.
\bibitem{Chakrabarti} C. G. Chakrabarti and I. Chakrabarty, {\it Mod. Phys. Lett.}, \textbf{20} (2006) 1471.
\end{thebibliography}
\end{document}